\documentclass[fleqn,twoside]{article}
\usepackage{espcrc2}

\usepackage{graphics}
\usepackage[figuresright]{rotating}

\def\ltsima{$\; \buildrel < \over \sim \;$}
\def\simlt{\lower.5ex\hbox{\ltsima}}
\def\gtsima{$\; \buildrel > \over \sim \;$}
\def\simgt{\lower.5ex\hbox{\gtsima}}
\def\ref{\par\noindent\hangindent 15pt}
\newcommand{\ea}{{et al.}}
\newcommand{\beq}{\begin{equation}}
\newcommand{\enq}{\end{equation}}
\newcommand{\bfg}{\begin{figure}}
\newcommand{\efg}{\end{figure}}
\newcommand{\bfa}{\begin{figure*}}
\newcommand{\efa}{\end{figure*}}

\newcommand{\AmS}{{\protect\the\textfont2
  A\kern-.1667em\lower.5ex\hbox{M}\kern-.125emS}}

\title{Iron Line Diagnostics for the GRS 1915+105 Black Hole}

\author{A. Martocchia\address[AMA]{ Dip. di Fisica "E. Amaldi", 
Terza Universit\`{a},  via della Vasca Navale 84, I--00146, Roma, Italia}
        \thanks{Present address: Observatoire Astronomique de Strasbourg,
        11 rue de l'Universit\'{e}Ž, -F-67000 Strasbourg, France. 
        Email: {\tt martok@isaac.u-strasbg.fr}},
        G. Matt\addressmark[AMA], 
        V. Karas\address{Astronomical Institute of the Charles University, 
	V Hole\v{s}ovi\v{c}k\'ach 2, CZ--180 00 Praha, Czech Republic}, 
        T. Belloni\address{Osservatorio Astronomico di Brera, via E.
	Bianchi 46, I--23807 Merate, Italy},
        M. Feroci\address{IAS/CNR, Area di Ricerca di Tor Vergata, Via
	Fosso del Cavaliere 100, I--00133 Roma, Italy}
}
       
\begin{document}

\begin{abstract}
The properties of the broad Fe line detected in two
{\it BeppoSAX} observations of the {\it microquasar} GRS 1915+105 are
summarized.
\vspace{1pc}
\end{abstract}

\maketitle

\section{INTRODUCTION}

Martocchia \ea\ (2002, {\bf Paper I} \cite{Mar02})  and 
Martocchia \ea\ (2003, {\bf Paper II} \cite{Mar03}) reported
the discovery of intense iron K$\alpha$ fluorescent emission
in two {\it BeppoSAX} observations of the {\it microquasar} GRS 1915+105,
which took place respectively on April 19-20, 1998, and April 21-22, 2000.
During these observations the Fe line is broad and asymmetric, 
best fitted with a relativistic disc model (see e.g. \cite{Fab89},  
\cite{Tan95}, \cite{Mar00} and Paper I for references on the relativistic
line model). 

These are the only two cases, to our knowledge, in which the 
iron K$\alpha$ line is so prominent in this source. 
The fact that the line is not always such strong
in GRS 1915+105 observations should not surprise, since finding 
the line out of the thermal emission high-energy tail
can be awkward in galactic Black Hole (BH) candidates, because of 
the high disk temperature. Nevertheless, strong, broad, possibly relativistic 
iron K$\alpha$ lines have recently been detected in various spectra 
of galactic BH candidates, e.g. Cyg X-1, XTE J1650-500,
XTE J1550-564, V4641 Sgr, GRO J1655-40, XTE J2012+381, GX 339-4,  
XTE J1748-288, XTE J1908+094.
They may help estimating the rotation parameter of the BHs hosted
at the center of these sources by means of the iron K$\alpha$
diagnostics.

\section{THE SOURCE}

GRS 1915+105 is a superluminal jet source, well-known 
since its discovery by the {\it WATCH} experiment on {\it Granat} 
 \cite{Cas92}. The source distance is about 10 kpc,
and its inclination -- based on observations of its jets -- 70 degrees. 
The optical counterpart was found only recently, via infrared observations, 
yielding evidence that GRS 1915+105 belongs to the class of 
low-mass X-ray binaries; the same observations allowed
to determine the mass of the central compact object, which 
has been constrained to $M_{\rm c}=14\pm4 M_\odot$, i.e. 
well above the standard neutron star mass limit (see \cite{Gre01}
and references therein). GRS 1915+105 is thus believed to host 
a BH with a gravitational radius $r_g=\frac{GM}{c^2} \sim 21$ km.

However, the issue of the central BH spin remains open. In the case of 
the two {\it BeppoSAX} observations here discussed, we could 
profit of the Fe line diagnostics, which allows to determine 
the accretion disk innermost stable orbit $r_{\rm ms}$, a known 
function of $J/M$ in Kerr metric, customarily 
put equal to the inner boundary of the optically thick portion 
of the disk itself.

GRS 1915+105 is a notoriously variable source, 
with the inner accretion disk being
blown-up and becoming optically thin for a relevant fraction of 
the time. This is one more reason for the line not to be always visible 
in {\it BeppoSAX} observations. We had to select appropriate time
intervals, out of both observations, in which the variability is less 
dramatic.

\begin{figure}
\includegraphics[angle=-90,width=0.45\textwidth]{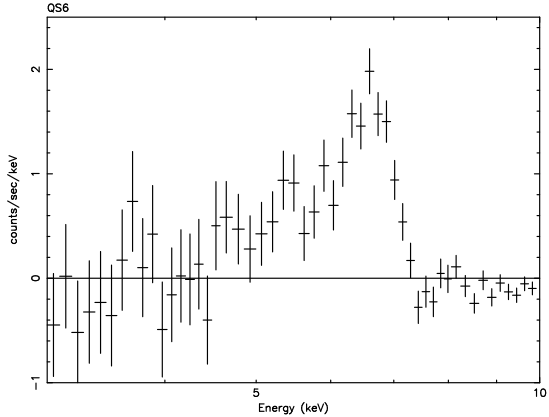}
\caption{Residuals at the iron line energies for one out of
six time intervals selected in the 1998 observation, with the
relativistic line model normalization put to zero for illustration
purposes (Paper I).}
\label{fig1}
\end{figure}

\section{DATA ANALYSIS}

Spectral fits have been performed with the {\sc XSPEC} software 
package. We assumed an optically thick, neutral, and geometrically 
thin Keplerian accretion disk with solar elemental abundances. 
For both data sets, results of a preliminary fit showed that the
dependence of the disc emissivity on the radius is rather flat: 
we assumed a phenomenological dependence $\sim r^{-2}$. 
``Cold'' absorption was also included: its column density parameter 
$n_{\rm H}$ always stabilizes at about $5.4$ and $5.45 \times 10^{22}$ 
cm$^{-1}$, in the two observations respectively; we thus choose 
these as fiducial values.

\subsection{First {\it BeppoSAX} observation: April 19-20, 1998 {\it (Paper I)}} 

This observation started at 11:33:19 UT and ended on the next
day at 20:14:52 UT. Out of the entire $\Delta t \sim 120$ ks, 
six time intervals of $\sim 2000$ s each were sorted out where the 
source variability is less pronounced. These are
six orbits in which the source is in a "quiescent" state. 
Data from the LECS, MECS and PDS instruments 
have been used in the intervals 0.1--4, 1.7--10 and 15--150 keV, 
respectively.  The basic ingredient to
model the continuum is a powerlaw with cutoff. 
As a model for the locally emitted
fluorescent and reflection features we initially considered 
numerically computed spectra,
including the Fe line and the underlying reflected continuum
derived from Monte Carlo
computations \cite{Mat91} and processed through our 
additive routine {\sc kerrspec}. 
However, we verified that the model is quite insensitive
to the Compton-reflected emission: to save computational
time we adopted for it a simple, non-relativistic model 
({\sc pexrav} in {\sc XSPEC}), and
used {\sc kerrspec} only to fit the line profile. 

We applied these ingredients on MECS and PDS data only.
The iron line parameters do not 
significantly change with more refined assumptions for the
continuum, for instance accounting for LECS data and including 
a multicolor blackbody component ({\sc diskpn}). Indeed,
the disc thermal emission is not clearly recognizable. 
Furthermore, the bestfit disk temperature comes out unrealistically 
high ($kT \sim 3.3$  keV), while the BH mass results far too small
from a physical point of view. These facts indicate that 
the standard interpretation of the continuum is not fully 
appropriate here. Here are the bestfit parameters:
$F_{2-10~{\rm keV}} \sim 1.8 \times 10^{-8}$ erg cm$^{-2}$ s$^{-1}$;
$\Gamma \sim 2.6 \div 2.7$; $E_{\rm cut} \sim 25$ keV.
The reflection parameter $R$ is not well constrained
and can vary from $\sim 0.15$ up to $1.2 \times 2\pi$. The innermost
line-emitting radius $r_{\rm in}$ stays always above $6r_{\rm g}$ (from
$\sim 7.8 r_{\rm g}$ in the fifth interval up to $\sim 79 r_{\rm g}$
in the second one); the Fe line equivalent width (EW) is
much larger than 100 eV in all cases, up to 300 eV in one interval.

Even though the $\chi^2$ statistics is not very good -- 
mainly due to the difficulty in modeling the continuum components --
the distorted profile (cp. Fig. 1) is always much better 
fitted with a relativistic line than with a
gaussian. We therefore infer that the relativistic solution 
for the iron line, besides being the most natural physical 
explanation for the very broad line observed, does not depend from
the choice of the continuum model. 

\begin{figure}
\includegraphics[angle=-90,width=0.45\textwidth]{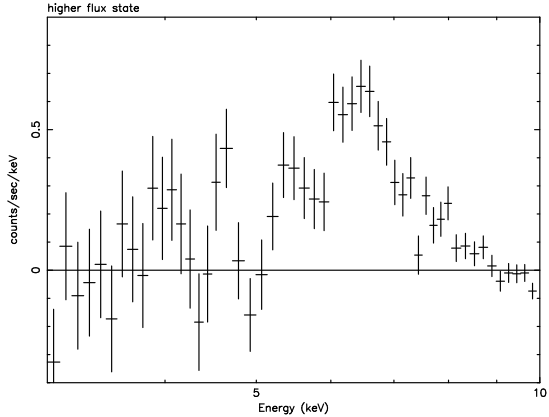}
\caption{Like in the previous Figure, but for one of the two
time intervals selected in the 2000 observation (Paper II).}
\label{fig2}
\end{figure}

\subsection{Second {\it BeppoSAX} observation: April 21-22, 2000 {\it (Paper II)}}

In this case we analyzed separately two selected --
``lower'' and ``higher flux'' -- spectral phases, lasting
$\sim 10$ ks each, out of the entire (19:56:55--11:16:55 UT)
observation.  Only data from the MECS and PDS instruments 
were available, and were used in the same energy intervals as above.
The flux is $F_{2-10~{\rm keV}} \sim 0.7$ and $1.1 \times 10^{-8}$ erg 
cm$^{-2}$ s$^{-1}$, in the two states respectively.
Again, we started our data analysis using a powerlaw 
with cutoff, disk multi-blackbody emission and Compton 
reflection ``hump''. Contrary to the 1998 
observation, thermal emission from the disk is not 
negligible this time. Broad and skewed residuals are apparent 
around 6.4 keV. 
The Fe feature is unphysically broad when modelled with a bare 
Gaussian ($\sigma \sim 2$ keV, $\chi^2_{\rm red} > 2$), suggesting 
relativistic broadening. We thus tried with a fully relativistic 
model  {\sc wabs * ( diskline + diskpn + refsch )} 
where {\sc refsch} accounts for relativistic distortions
in the reflected continuum. Since on this way 
the best-fit innermost line-emitting orbit always results to 
be equal to the inner hard limit -- $6 r_{\rm g}$, the last stable 
orbit in Schwarzschild metric -- it has been straightforward 
to test the {\it spinning} BH assumption with these data. 
We employed the Kerr line model {\sc kerrspec} (cp. Paper I)
to replace {\sc diskline}. Best-fit parameters are shown in the
Table. The iron line rest energy is $E_0 \equiv 6.4$ keV. 
The value of the farthest emitting radius $r_{\rm out}$ is not 
well constrained by the iron line models, and always tends 
to the outer hard limit set by the model itself. 

\begin{table}[htb]
\caption{Bestfit parameters values for the 2000 observation (Paper II).
The errors refer to $\Delta \chi^2 = 2.706$. 
For comparison, the $\chi^2/dof$ for the case $r_{\rm in}/r_{\rm g} 
\equiv 6$ (i.e. in a static BH assumption) is also shown (at bottom). 
Even if the improvement in the $\chi^2$ statistics is not very
significant, 
remarkably $r_{\rm in}$ stays always below $6 r_{\rm g}$ 
when adopting a (canonical, $J/M = 0.9981$) spinning BH model. 
No significant changes of the physical 
parameters can be appreciated when the 1\% systematics is employed.}
\vskip2pt
\newcommand{\m}{\hphantom{$-$}}
\newcommand{\cc}[1]{\multicolumn{1}{c}{#1}}
\renewcommand{\arraystretch}{1.2} 
\hskip-4mm
\begin{tabular}{@{}lll}
\hline
& & \cr
       & {\it ``Lower flux''} & {\it ``Higher flux''} \cr
& {\it sp. ``state''} & {\it sp. ``state''} \cr
& & \cr
\hline
& & \cr
$\Gamma$            & $2.17^{+0.05}_{-0.02}$ & $2.59^{+0.04}_{-0.04}$ \cr
& & \cr
$E_{\rm cut}$ [keV] & $78.8^{+3.9}_{-4.0}$   & $183.7^{+4.1}_{-3.8}$      \cr
& & \cr
$R$                 & $0.92^{+0.22}_{-0.09}$ & $1.72^{+0.14}_{-0.03}$   \cr
& & \cr
$\xi$ (reflection) [erg ${{\rm cm}\over{\rm s}}$] & $10.3^{+47.0}_{-8.6}$ 
                                     & $0.000^{+0.001}_{-0.000}$ \cr
& & \cr
$T$ (reflection) [eV] & $28.7^{+61.0}_{-10.2}$ & $85.1^{+49.3}_{-21.0}$ \cr
& & \cr
$T$ (blackbody) [eV] & $393^{+41}_{-37}$ & $1414^{+63}_{-19}$ \cr
& & \cr
$r_{\rm in}/r_{\rm g}$ (blackbody) 
                     & $10.0^{+0.6}_{-0.7}$ & $79.2^{+0.9}_{-1.4}$ \cr
& & \cr
\hline
& & \cr
$r_{\rm in}/r_g$ (Fe K$\alpha$) 
       & $3.9^{+16.6}_{-1.6}$ & $1.4^{+3.7}_{-0.2}$  \cr 
& & \cr
Fe K$\alpha$ EW [eV]     & 125 & 126 \cr
& & \cr
$\chi^2/dof$                         & $148.2/86$ & $144.9/86$ \cr
& & \cr
\hline
& & \cr
$\chi^2/dof$ ($r_{\rm in}/r_{\rm g} \equiv 6$) & $148.9/86$ & $154.9/86$ \cr
& & \cr
\hline
\end{tabular}
\label{tab:kerr}
\end{table}

\section{DISCUSSION}

The Fe K$\alpha$ fluorescent emission feature detected in both
1998 and 2000 {\it BeppoSAX} observations of the {\it microquasar}
GRS 1915+105 is strong, broad and asymmetric, best-fitted
with a relativistic disc model. The line is emitted from neutral 
or low ionized iron. In the case of the 1998 observation 
we found evidence of emission from a region of the disk 
which is near, but still outside the Schwarschild innermost stable 
orbit; on the other hand, in the 2000 spectrum emission 
from $r<6r_{\rm g}$ (the last stable orbit in Schwarzschild metric) is
compatible with the data, even if the results of our data analysis do 
not allow a firm conclusion on this regard. Emission from inside 
$6r_{\rm g}$ would indicate a Kerr spacetime, 
i.e. that a rotating BH is hosted at the center of the system.
Of course, the disc parameters are 
somewhat dependent on the adopted model for the
continuum, and must be taken with caution. Since the BH spin stays 
constant between the two observations, differences in the 
innermost emitting radii must reflect changes in the accretion 
flow.

Most of the observed broad, intense Fe lines in 
galactic BH candidates have beed detected during ``very high'', 
``intermediate'' or ``quiescent'' spectral states. A plausible 
explanation for this has still to be found: to produce the line,
the accretion disc must efficiently reflect primary photons, even 
at the innermost 
radii. This is unlikely in ``hard'' states, since such states 
correspond to absent or optically-thin disks, but could be 
the case of ``soft/high'' states. 

In the GRS 1915+105 1998 observation we registered 
very strong flux and an extremely steep powerlaw, with a cutoff at 
rather low energies. In the 2000 observation 
we have a similar powerlaw continuum, but thermal emission
is clearly seen, and ``cold'' reflection is better constrained.
Both observations may correspond to ``very high'' states, with
Comptonization tails originating from the disk blackbody seed photons. 
BH mass estimates 
inferred using the thermal luminosity are physically unplausible,
also due to the fact that current thermal emission models 
are not adequate to describe radiation transfer very near to the 
event horizon.

More detailed discussions on the data analysis and
results are reported in Paper I and Paper II. Various issues 
still remain open, with regard to the accretion
disk geometrical and physical states. Data with enhanced spectral 
resolution and sensitivity are strongly needed, and will be hopefully 
achieved by the next observational campaigns, first of all
the ones scheduled for the satellite {\it XMM-Newton}.

\end{document}